\documentclass[journal=jacsat,manuscript=communication,layout=twocolumn,articletitle=true]{achemso}

\usepackage[version=3]{mhchem} 
\usepackage{graphicx}
\usepackage[T1]{fontenc}       
\usepackage{dblfloatfix}
\usepackage{gensymb}
\usepackage{grffile}
\usepackage{chngcntr}
\usepackage{fixltx2e}
\usepackage{chemformula}
\usepackage[T1]{fontenc} 
\usepackage[numbers]{natbib}
\usepackage{subfig}
\usepackage{epsfig}
\usepackage{braket}
\usepackage{psfrag}
\usepackage{subfig}
\usepackage{multirow}
\usepackage{xr}
\usepackage{mathtools}
\usepackage{amsmath}
\usepackage{amssymb}
\usepackage{commath}
\usepackage{amsfonts}
\usepackage[normalem]{ulem}
\usepackage{braket}
\usepackage{array}
\usepackage{tabularx}
\usepackage{stackengine}

\setkeys{acs}{articletitle = true}
\SectionsOn

\author{\'Angel Fern\'andez-Blanco}
\affiliation{Institut Laue Langevin, 71 Avenue des Martyrs, CS 20156-38042, Grenoble, France}  
\alsoaffiliation{SIMaP, University of Grenoble-Alpes, 38042 Grenoble, France}
\author{Luc\'ia Pi\~{n}eiro-L\'opez}
\affiliation{IMDEA Nanociencia, Faraday 9, Ciudad Universitaria de Cantoblanco, 28049 Madrid, Spain}
\author{M\'onica Jim\'enez-Ruiz}
\affiliation{Institut Laue Langevin, 71 Avenue des Martyrs, CS 20156-38042, Grenoble, France}  
\author{Stephane Rols}
\affiliation{Institut Laue Langevin, 71 Avenue des Martyrs, CS 20156-38042, Grenoble, France}  
\author{Jos\'e Antonio Real}
\affiliation{Departamento de Qu\'imica Inorg\'anica, Insituto de Ciencia Molecular (ICMol), Universidad de Valencia, Spain}
\author{J. Alberto Rodr\'iguez-Velamaz\'an}
\email{velamazan@ill.eu}
\affiliation{Institut Laue Langevin, 71 Avenue des Martyrs, CS 20156-38042, Grenoble, France}  
\author{Roberta Poloni}
\email{roberta.poloni@grenoble-inp.fr}
\affiliation{Grenoble-INP, SIMaP, University of Grenoble-Alpes, CNRS, 38042 Grenoble, France}

\title[An \textsf{achemso} demo]{Probing the SO$_2$ adsorption mechanism in Hofmann clathrates via inelastic neutron scattering and density functional theory calculations}

\abbreviations{}

\begin{document}
\maketitle

\begin{abstract}
  The adsorption mechanism of SO$_2$ in the Hofmann-like coordination polymer {Fe(pz)[Pt(CN)$_4$]} is studied using inelastic neutron scattering and density functional theory calculations.  We find that the most important spectral change upon gas adsorption is the blueshift of the low energy peak found at 100 cm$^{-1}$, a feature that is fully confirmed by the computed neutron-weighted phonon density of states. Our calculations suggest that the origin of this change is twofold: i) an increase in the force constant of the cyanide out-of-plane movement due to the binding of the gas onto the Pt(CN)$_4$ plane, and ii) the hampered rotation of the pyrazine due to steric hindrance. The high energy region of the neutron scattering data whose spectral weight is dominated by the internal vibrations of the pyrazine is negligibly affected by the presence of the gas as expected from a physisorption type of binding.

\end{abstract}

\section{Introduction}
Metal-organic frameworks (MOFs) are 3D nanoporous materials formed through coordination bonds between metal cations and organic ligands.  The great variety of metal ions, organic linkers, combined with the structural topology, allow to achieve an almost infinite number of possible combinations.
 Owing to this exceptional tunability, combined with the larger surface area and nanoscale porosity, the past 10-15 years have seen a rapid development in the field of MOFs for efficient gas adsorption and separation \cite{B802426J,Yang2018}.
 In recent years, adsorption and chemical sensing of toxic gas molecules using MOFs has become a very active field of research \cite{He2013,sensing,Gamonal2020}. Among these, iron (II) spin-crossover Hofmann-type clathrates represent an interesting class of MOFs with sensing capabilities due to their bistability, meaning that they can be {\sl switched} between two different spin states \cite{Munoz2011,Ni2017}. These complexes undergo a spin-state change under the influence of external stimuli such as light, temperature, pressure or the incorporation of guest molecules.
 A relevant feature is the presence of metallic centers with an ``open'' metal coordination, i.e. metal centers exhibiting an unsaturated coordination. Interestingly, these open-metal sites have been shown to exhibit a high affinity for many gases, such as CO$_2$\cite{Yaghi2019}, CH$_4$ \cite{Guo2011}, and CO\cite{Reed2017}.
 Based on the well known Hofmann clathrate compounds~\cite{Hofmann1897}, these materials are built via cyanide (CN) bridging ligands forming metallo-cyanide planes with different linkers acting as bridges between the planes. The most representative example of this class of compounds is the family {Fe(pz)[M(CN)$_4$]}, with M = Ni, Pd, Pt, and pz = pyrazine\cite{Niel2001}. The Fe(II) centers undergo a transition (spin crossover) from low spin to high spin, yielding a change in the magnetic, optical, dielectric, and structural properties of the material.   This bistability coupled with the presence of potentially high-affinity open metal sites makes them excellent candidate materials for  MOF-based gas sensing switches \cite{sensing}.
 \par In 2013, Arc\'is-Castillo et {\sl al.}~\cite{Real2013} studied the adsorption of SO$_2$ in the {Fe(pz)[Pt(CN)$_4$]} Hofmann-type clathrate by measuring the adsorption isotherms and the x-ray diffraction patterns. The experimental results, combined with the DFT calculations established that the SO$_2$ binds strongly via chemisorption. In this work, we combine inelastic neutron scattering data with DFT calculations to further understand and clarify the SO$_2$ adsorption mechanism in the material. The most relevant signature of the binding occurs in the spectral region between 100 cm$^{-1}$ and 140 cm$^{-1}$. The intense low energy peak at around 100 cm$^{-1}$ blueshitfs upon adsorption, a signature that we attribute to both the hampered rotation of the pyrazine and the out-of-plane movement of the cyanide.
 The well defined peaks in the high-energy region of the spectra are associated with the internal vibrations of the pyrazine and are negligibly affected by the adsorbed gas. These findings are well reproduced by the computed generalized-phonon density of states and are consistent with the analysis of a strong physisorption mechanism occurring via the Pt(CN)$_4$ plane and the pyrazine molecules.  This work shows that INS measurements provide a powerful tool to probe the gas adsorption mechanism in this class of materials thus allowing to unambiguosly characterize the nature of the interaction when supported by DFT calculations.
 
\section{Methods}
\textbf{Sample preparation.} [Fe(pz)Pt(CN)$_4$]·nH$_2$O precipitates when a solution of K$_2$[Pt(CN)$_4$] in H$_2$O is added with constant stirring to a solution which contains stoichiometric amounts of pz and Fe(BF$_4$)$_2$·6H$_2$O in MeOH/H$_2$O (1:1) under nitrogen at room temperature. A small amount of ascorbic acid was added to prevent the oxidation of Fe(II). After stirring for 1h, the yellow precipitate was collected by suction filtration, washed with water and methanol and dried under ambient pressure \cite{Niel2001}. The sample was then heated in a drying oven at a temperature of 100 $\degree$C. \\

\textbf{Gas adsorption.} The sample (722.9 mg, 1.6 mmol) was placed inside a cylindrical aluminium sample holder allowing the gas injection, connected with a capilar to a manifold gas pumping system. Temperature control was achieved using either a closed cycle cryostat (IN1-LAGRANGE) or an Orange cryostat ({\sc IN5} and PANTHER). The empty MOF was measured at 30 K. The adsorption of SO$_2$ was performed near room temperature in two injection steps until reaching a total of 1.6 mmol of SO$_2$ adsorbed, that is, one molecule of SO$_2$ per formula unit of the host. Finally, the temperature of the sample was decreased to 30 K to perform the measurements.  \\

\textbf{IN1-LAGRANGE.} The INS experiment was performed in the indirect geometry-type spectrometer IN1-LAGRANGE \cite{Ivanov2014} installed on the hot neutron source of the high flux reactor at the Institut Laue-Langevin (ILL) in Grenoble, France. The incident neutron energy was determined using a combination of Cu and Si monochromators for the intermediate and lower energy range respectively. Upon scattering by the sample, neutrons enter a secondary spectrometer where a beryllium filter is installed to remove higher-order harmonics in the analyser reflections. Neutrons are then reflected to a He3 gas detector at fixed final energy of 4.5 meV. This is done by using a focussing analyzer built around the vertical sample-detector axis.
The scattering angle, $\theta$, varies from 33.7$^{\circ}$ to 69.4$^{\circ}$ and the accesible kinematical range of the instrument, $Q$, is defined as

\begin{equation}
  Q^2=k_i^2 + k_f^2 - 2k_ik_fcos(\theta)
  \label{eq1}
\end{equation}

Where k$_i$ ad k$_f$ are the initial and final scattering wave vectors that can be related to neutron energy by E=$\frac{\hbar^2k^2}{2m_n}$, m$_n$ being the neutron mass.  To collect the data, the incident neutron energy was scanned step-by-step by rotating the whole secondary spectrometer around the monochromator allowing the measurement of a neutron energy loss spectra. The monochromators were selected to provide the best relation between energy transfer range and resolution: we employed Cu(220), Si(311), and Si(111), respectively, for the energy transfers of [44 - 161] cm$^{-1}$, [130 - 286] cm$^{-1}$, and [209 - 1500] cm$^{-1}$. Data were collected at 30 K for both the empty and loaded material.
\\

\textbf{IN5.} INS data for the bare material were collected in the low energy region, i.e.~$<$ 100 cm$^{-1}$, as a function of momentum transfer on the direct geometry disk chopper time-of-flight spectrometer at the ILL (Grenoble, France). The cold neutron flux from the neutron guide H16 is scattered by the sample after being turned into a pulsed monochromatic beam by a set of 6 choppers. The scattered neutrons are collected by a large cylindrical array of pixelated position-sensitive detectors (PSD) mounted inside a vacuum time-of-flight chamber, giving access to the dynamic structure factor $S(Q,\omega$). Data were collected at 295 K (at low-spin state) using the standard configuration with a wavelength of 5 \AA\ (E$_i$ $\approx$ 29.8 cm$^{-1}$) and a Q range of $\backsim$ 0.2 - 4.1 \AA$^{-1}$ for the angular detector coverage.\\

\textbf{PANTHER.} Complementary measurements for the low energy region were performed in the high-flux direct-geometry time-of-flight spectrometer PANTHER installed on the H12 thermal beam tube at the ILL. A double focusing pyrolytic graphite monochromator determines the incoming energy in the range 7.5 - 112 meV. Short neutron pulses produced by a slit Fermi chopper are scattered by the sample and collected by PSD. Data were collected at 10 K using an initial energy of 50 meV. \\

\textbf{Computational details} The DFT calculations were performed with the Quantum Espresso package \cite{Giannozzi2009,Giannozzi2017,Giannozzi2020} (v.~6.4) within the generalized gradient approximation (GGA) of Perdew, Burke and Ernzerhof (PBE)\cite{Perdew1996} and long-range interactions described with the semiempirical approach proposed by Grimme (PBE+D2)\cite{Grimme2006,Grimme2016}. We use the Rappe-Rabe-Kaxiras-Joannopoulos ultrasoft (rrkjus) pseudopotentials \cite{Rappe1990} without semicore states in valence. The convergence threshold on forces is 0.0001 Ry/Bohr and the wavefunctions and charge density cutoffs are set to 100 Ry and 1000 Ry, respectively. These are carefully chosen to obtain converged phonon frequencies.
The low temperature (low spin) structure of the Hofmann clathrate was described by many authors as a disordered orientation of the pyrazines \cite{Velamazan2012,Niel2001,Kepert2009,Ohba2009,Cobo2008}. Recently, the present authors employed neutron diffraction data collected at D20@ILL to show an ordered configuration with the pyrazines perpendicular to each other \cite{Fernandez-Blanco2021}, a configuration that has already been observed in certain conditions~\cite{Kepert2009,Ohba2009,RODRIGUEZ2010,Aravena2014}. To describe this perpendicular orientation a supercell with lattice parameter $a'$=$\sqrt{2}a$ and $b'$=$\sqrt{2}b$ was used, with $a$ and $b$ lattice parameters of the primitive cell. This impos implying a disordered  orientation (parallel in the calculations) of the pyrazines.  The PBE+D2 lattice parameters of the bare MOF are i.e. $a$ = 10.096 \AA, $b$ = 10.097 \AA\ and $c$ = 6.711 \AA. The Brioullin zone is sampled using 3$\times$3$\times$3 Monkhorst-Pack k-points grid. A revised PBE approximation for densely packed solids by Perdew et. al \cite{Perdew2008} (PBEsol+D2) and the nonlocal functional rev-vdW-DF2 \cite{Hamada2014} were used to study the effect of the functional choice on the computed density of states (see text). For rev-vdW-DF2 we employ PBE-generated pseudopotentials while for PBEsol+D2, since the convergence of the electronic structure  with the rrkjus was not reached, we employed pseudopotentials from Garrity-Bennet-Rabe-Vanderbilt library (GBRV) \cite{Garrity2014}. Wavefunctions and charge density cutooffs used for PBEsol+D2 and rev-vdW-DF2 are 70 Ry and 700 Ry. Calculations are performed using the low spin (S=0) electronic configuration.
\\

\begin{figure*}[ht]
  \centering
  \includegraphics[scale=0.6]{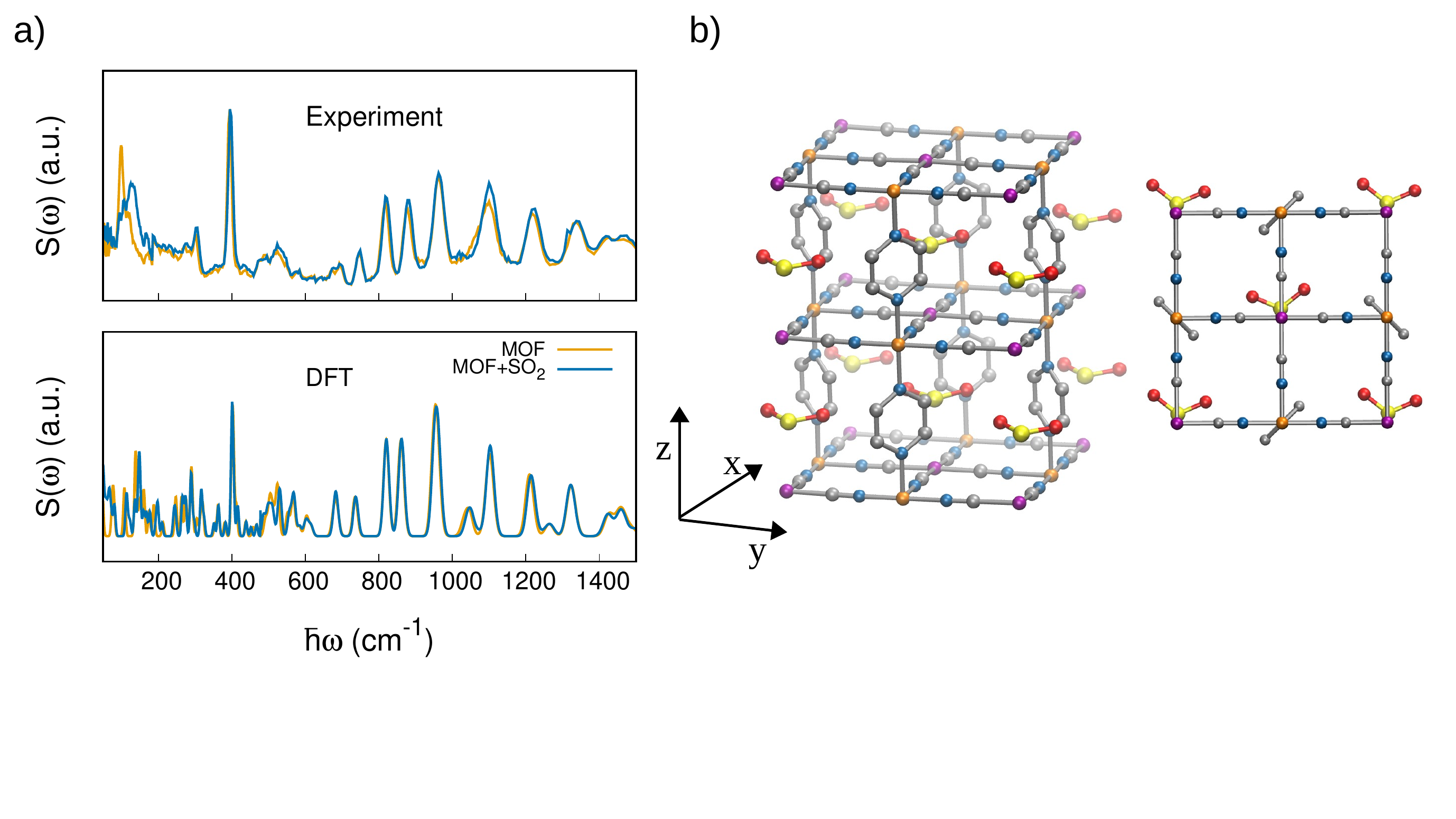}
  \vspace{-1.5cm}
  \caption{Experimental inelastic neutron scattering data, S($\omega$), collected at 30 K at IN1-LAGRANGE for the bare compound and upon SO$_2$ uptake, upper panel (a). The S($\omega$) computed using the PBE+D2 functional is reported in the lower panel (a). Side and top view illustrations of unit cell employed for the calculations containing neighboring pyrazine molecules oriented perpendicularly. Color code: purple, orange, blue, silver, red, and yellow are Pt, Fe, N, C, O, and S, respectively. H atoms are omitted for clarity.}
  \label{Fig1}
\end{figure*}

\textbf{Generalized-phonon density of states.} The phonon frequencies were obtained by diagonalizing the dynamical matrix by employing the ph.x package in Quantum Espresso\cite{Giannozzi2009}. The harmonic interatomic force constants are computed using density functional perturbation theory \cite{Baroni2001,Gonze1997}. The inelastic scattering data collected at IN1-LAGRANGE are compared with the computed generalized-, or neutron-weighted, phonon density of states. The $l$-atom contribution to the total one-phonon density of states is given by
\begin{equation}
  g_l(\omega)=\sum_{q,j} |u_{l}(j,\textbf{Q})|^2 \delta(\omega-\omega_j(\textbf{Q}))
  \label{eq3}
\end{equation}
where $\omega_j(\textbf{Q})$ is the phonon frequency of the $j$-branch at $\textbf{Q}$ wavevector and $u_{l}(j,\textbf{Q})$ is the corresponding atomic displacement. Thus, $g(\omega)d\omega$ gives the number of eigenstates in the frequency interval ($\omega$, $\omega$+$d\omega$).
The total generalized-phonon density of states, G($\omega$), is then defined as the phonon density of states weighted by the neutron scattering power of each atom $l$ and it is given by
\begin{equation}
  G(\omega)=\sum_{l=1}g_{l}(\omega)\frac{\sigma_{l}}{M_l}
  \label{eq4}
\end{equation}
where $\sigma_{l}$ is the incoherent neutron scattering cross section and $M_l$ is the mass. 
In the harmonic and incoherent approximation, this density of vibrational states is related to scattering function $S(Q,\omega)$ by the following relation\cite{Squires1996,Carpenter1987,Elliott1997}
\begin{equation}
  S(Q,\omega) = e^{-2\overline{W}}\frac{Q^2\hbar}{2\overline{M}\omega}<n+1>G(\omega)
  \label{eq5}
\end{equation}
where $\overline{M}$=$\sum_l$ M$_l$/N, $n$ is the thermal-equilibrium occupation number of the vibrational state and $<n+1>=\frac{exp(\hbar\omega\beta)}{exp(\hbar\omega\beta)-1}$ and $\beta=\frac{1}{K_BT}$. We drop the vector symbol from  the scattering function because for powders we measure the average over momentum transfer. The exponential term is the Debye-Waller factor for neutron attenuation by thermal motion and 2W=$\frac{Q^2\bar{<u^2>}}{3}$. The average of the mean-square displacements over all the atoms is $<\overline{u^2}>$ and it is computed as
\begin{equation}
  <\overline{u^2}> = \frac{\hbar}{2\overline{M}N}\sum\frac{1}{\omega_j}coth(\frac{1}{2}\hbar\omega_j\beta)
  \label{eq6}
\end{equation}
where $Q$ is the kinematical range of IN1-LAGRANGE (eq.~\ref{eq1}).
Because of the small $Q$ range measured at each energy transfer, we compute phonons only at $\Gamma$ to compare with experimental INS data, thus dropping the sum over the phonon wavevector in eq.~\ref{eq3}. Thus eq.~\ref{eq5} becomes $S(\omega)$. This should be a fair approximation since we are measuring small $Q$ at low energy while at high energy optical modes show small dispersion.
Finally, the phonon density of states is convoluted with a Gaussian function to account for the resolution of the monochromators. We set the standard deviation of the Gaussian to 3.0 cm$^{-1}$ for the range [0 - 478] cm$^{-1}$ and 0.009$\omega$ for [478 - 4033] cm$^{-1}$, close to the experimental resolution of IN1-LAGRANGE~\cite{Ivanov2014}.


\section{Results and discussion}
\begin{figure}[ht!]
  \centering
\includegraphics[scale=0.38]{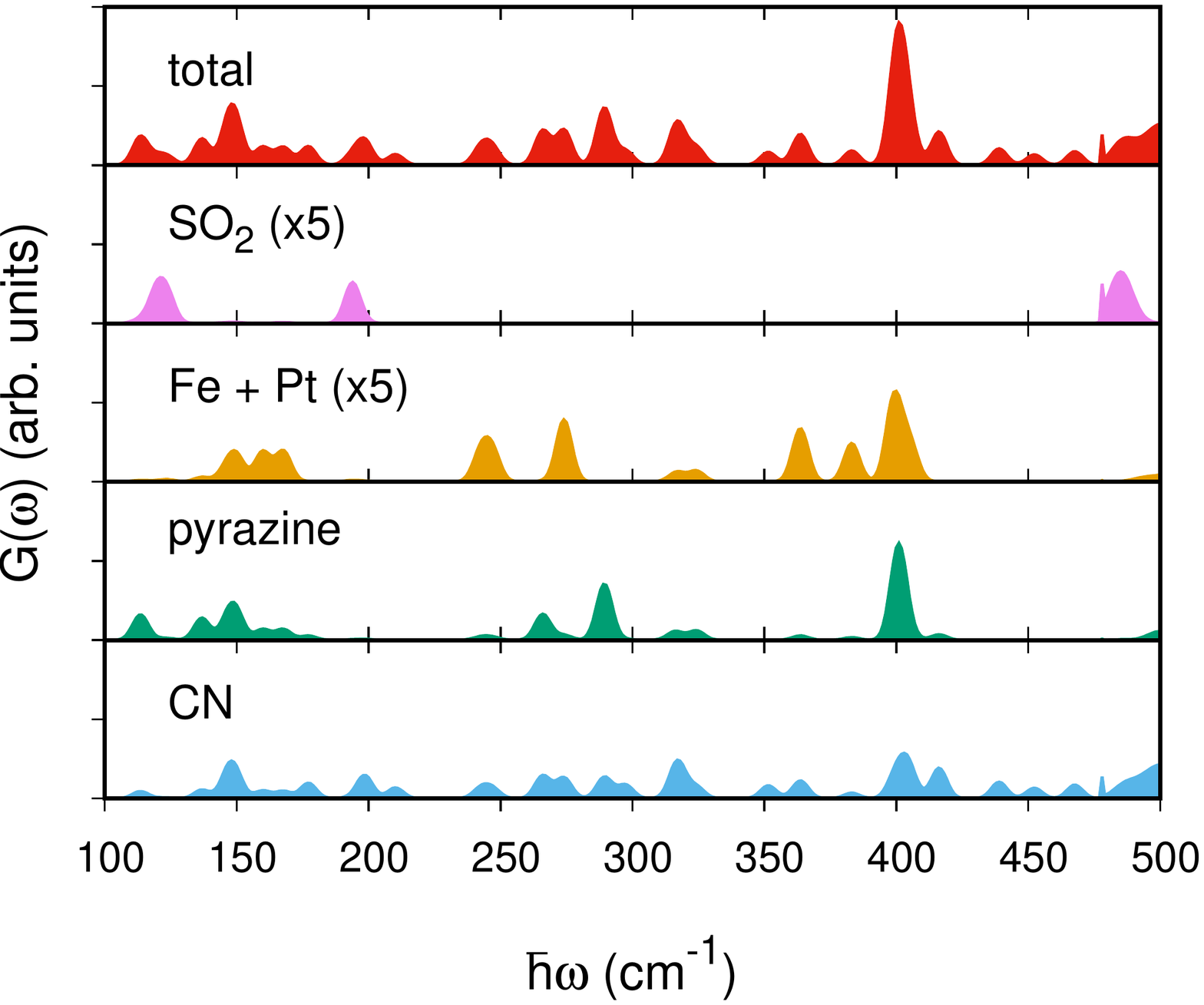}
\includegraphics[scale=0.38]{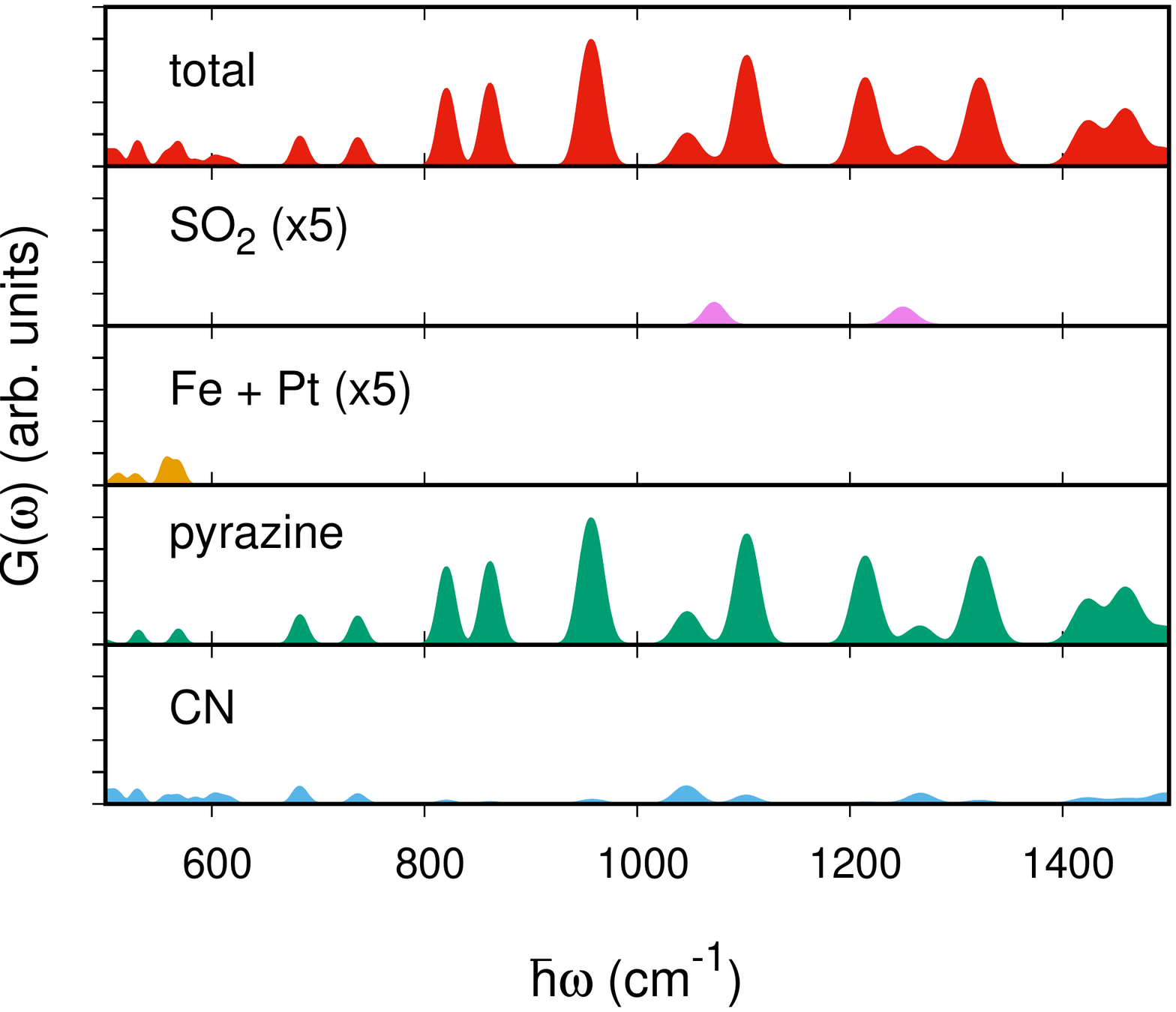}
\vspace*{-1.5cm}
\caption{Total and partial (atom-specific) generalized phonon density of states computed for the Hofmann clathrate with adsorbed SO$_2$ molecules. The low energy part and the high energy part are plotted separately in upper and lower panels, respectively. The partial g($\omega$) of SO$_2$ and Pt+Fe is multiplied by a factor of 5 in order to be visualized in the same scale of the $y$ axis.}
  \label{Fig2}
\end{figure}

The INS spectra collected at IN1-LAGRANGE are shown in the upper panel of Fig.~\ref{Fig1} for the empty and loaded MOF. The scattering function computed using eq.~\ref{eq5} is shown in the lower panel of Fig.~\ref{Fig1}. The full spectra including experimental errors are reported in Fig.~S1. 
We note the need to rescale the computed $<\overline{u^2}>$ by a factor of 10 in order to reproduce the experimental energy decay of the spectra. This deviation may derive from the fact that only the normal modes are employed to compute eq.~\ref{eq6} which is a rough approximation to the full density of vibrational states, especially in the low energy region which dominates $<\overline{u^2}>$.
\par Since the IN1 measurement is not optimized below 100 cm$^{-1}$ (see Fig.~S1), we studied this energy region for the bare material by using the the time-of-flight spectrometer IN5. The data are reported in Fig.~S2. These low-energy vibrational modes are associated with the vibrations of the whole lattice and with out-of-plane and in-plane vibrations of the Pt atom. Because of the difference between the computed normal modes at $\Gamma$ and the experimental data, the assignement of the modes within this region is somehow problematic.
\par Between 75 and 450 cm$^{-1}$ of the measured INS data, two main peaks are observed for the bare material, at 100 cm$^{-1}$ and 393 cm$^{-1}$, and in between these two a less resolved region appears. Upon adsorption of SO$_2$ the peak at 100 cm$^{-1}$ becomes broader and shifts to higher energy, the maximum being found at 129 cm$^{-1}$. The remaining part of the spectrum is negligibly affected by the presence of the gas, even the strong excitation at 393 cm$^{-1}$. This behavior is in excellent  agreement with the computational results.
\par The high energy region between 700  and 1500 cm$^{-1}$ reveals several well defined excitations which are very well reproduced by the simulations and whose nature will be discussed later.  The negligible change upon gas adsorption measured in this region is also confirmed by the calculations. 

\subsection{Partial G($\omega$)}

To better analyze the nature of the measured excitations we report in Fig.~\ref{Fig2} the partial G($\omega$), i.e. the $g_{l}(\omega)\frac{\sigma_{l}}{M_l}$, where $l$ represents the different atoms or molecules in the MOF, for the Hofmann clathrate with adsorbed SO$_2$ molecules.

\begin{figure}[ht]
  \centering
  \includegraphics[scale=0.36]{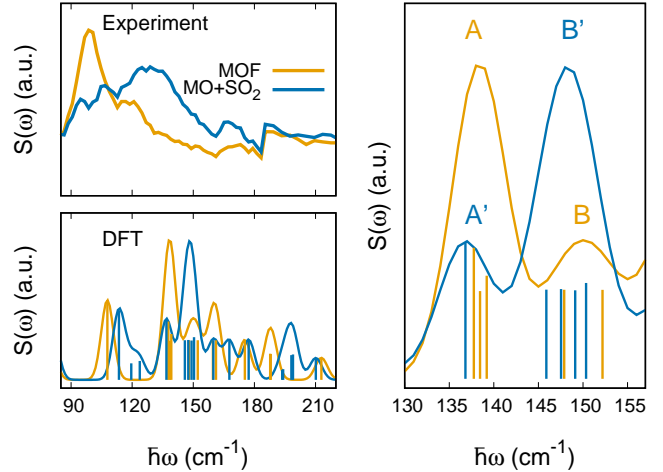}
  \caption{Left panels: INS spectra measured at IN1 at 30 K for the empty Fe(pz)[Pt(CN)$_4$] (orange) and after SO$_2$ adsorption (blue) in the upper figure. The lower figure reports the computed S($\omega$). Rigth panel: zoom in the region [130 - 160] cm$^{-1}$ where the main change upon gas adsoprtion occurs. }
  \label{Fig4}
\end{figure}

In the low energy region, between 20 and 100 cm$^{-1}$, the G($\omega$) is dominated by the vibrations of the heavy atoms, Fe and Pt, and to a minor extent by vibrations of the pyrazines and the CN groups. A few librational modes of SO$_2$ also appear between 20 and 120 cm$^{-1}$  (see Fig.~S3). Concerning the SO$_2$ molecule, the normal modes in the gas phase are predicted at 484 cm$^{-1}$, 1107.7 cm$^{-1}$ and 1303.5 cm$^{-1}$ (see Fig.~S3). These are respectively the scissoring (i.e.~the asymmetric bending), and the symmetric and asymmetric stretches. All these modes are IR active and are found to change negligibly upon SO$_2$ adsorption (see Fig.~S3). When the molecule is adsorbed in the MOF, these are computed at  484.5 and 485.6 cm$^{-1}$, 1042.2 and 1072.4 cm$^{-1}$, and 1248.9 and 1250.4 cm$^{-1}$ (2 molecules per unit cell). 
\par Between 100 and 500 cm$^{-1}$, all atoms contribute to the generalized density of states as shown in the upper panel of Fig.~\ref{Fig2}. The peaks predicted at $ca.$ 140 cm$^{-1}$ and 400 cm$^{-1}$ have contributions mainly from the pyrazine and the CN groups as explained in more detail below, while the peak computed at 107 cm$^{-1}$ is mainly due to a vibration of the pyrazine.
At high energy, between 600 and 1500 cm$^{-1}$, the heavy atoms do not contribute any longer and the CN contribution is almost negligible. Here the well defined peaks are associated with vibrations of the pyrazine and their assignement is reported later. These bands exhibit a strong spectral weight due to the contribution from H atoms.

\subsection{Low-energy region}
The main change upon adsorption is found in the peak at 100 cm$^{-1}$ (see Fig.~\ref{Fig1}). The intensity of this peak decreases and a new broad feature appears centered at about 129 cm$^{-1}$. This result is confirmed by the INS data collected on PANTHER which are shown in Fig.~S4. The peak measured on PANTHER at 96.7 cm$^{-1}$ for the bare material should correspond to the intense band measured at 94.4 cm-1 on IN1 (see Fig.S2).
In this region, the most intense bands predicted by the calculations for the bare material appear at 107.9 cm$^{-1}$ and at 139 cm$^{-1}$, as shown in Fig.~\ref{Fig4}. The first one corresponds to the rotation of the pyrazine around the $z$ axis, in agreement with calculations by Hochd\"{o}rffer et al.~\cite{Hoch2019} on a 3D molecular cluster composed of several repetitions of the clathrate unit cell. The second band includes 3 vibrational modes at energies 137.8, 138.5, and 139.2 cm$^{-1}$. We name this second intense peak A in Fig.~\ref{Fig4} to assist the analysis. The first peak at 137.8 is a collective mode involving a movement of the CN group with contributions from in-plane and out-of-plane, together with a rotation of the pyrazine around the $z$ axis. The second at 138.5 cm$^{-1}$ involves a rigid out-of-plane twisting of the Fe(CN$_4$)N$_2$ octahedra yielding a rigid twisting of the pyrazine about the $y$ axis. This vibrational mode correponds to the one computed at 140 cm$^{-1}$ in ref.~\citenum{Hoch2019}. The vibration at 139.2 cm$^{-1}$ involves a large out-of-plane vibration of the cyanide together with a small libration of the pyrazine.  The change in the eigenvectors is illustrated in Fig.~S5.
Upon gas uptake, the vibration at 107.9 cm$^{-1}$ blueshits to 113.5 cm$^{-1}$ possibly due to steric hindrance by the SO$_2$ molecule and it becomes coupled with a  libration of SO$_2$. Two new modes consisting of pure SO$_2$ librations appear at 119.5 and 123.7 cm$^{-1}$.
Under peak A, the first mode barely changes (i.e.~137.8 $\to$ 136.8  cm$^{-1}$) while the second and third blueshift by 9 and 10  cm$^{-1}$, respectively (i.e.~138.5 $\to$ 147.5 cm$^{-1}$, and 139.2 $\to$ 149.1 cm$^{-1}$). The Pt atom does not participate in these vibrations for the bare MOF, but upon gas adsorption the heavy atom contributes to the out-of-plane bending of the cyanide for the mode at 149.1 cm$^{-1}$.  For the bare MOF the peak named B in Fig.~\ref{Fig4} includes two vibrations at 147.9 cm$^{-1}$ and 152.2  cm$^{-1}$.   These correspond to an in-plane movement of the Pt atoms together with a rigid in-plane displacement of the pyrazines. Upon adsorption these slightly redshift to 145.9 cm$^{-1}$ and 150.4  cm$^{-1}$, respectively. All of the vibrations found in peaks A$^\prime$ and B$^\prime$ exhibit a negligible contribution from SO$_2$.

Because the scattering cross section of hydrogen, $\sigma_H$ = 82.0 barn, is significantly larger than any other atom (Fe and Pt for example have 11.62 and 11.71 barn, respectively), the gas adsorption mechanism is here probed mainly through changes in the riding modes of H, i.e.~those modes that involve vibrations of the hydrogens \cite{MitchellVibrational}. 
As described above, in this region, the vibrations that imply the rotation of the pyrazine around the $z$ axis and the cyanide out-of-plane movement shift to higher energy while those exhibiting in-plane bendings of cyanides show a negligible shift.  We attribute the blueshift of this peak predicted at $ca.$ 139 cm$^{-1}$ to an increase of the cyanide out-of-plane bending force constant due to steric hindrance, similarly to the hindered rotation of the pyrazine predicted at 107 cm$^{-1}$.
\par While the assignement of the experimental peak at 400 cm$^{-1}$ can be performed without ambiguity based on the good agreement with the calculations, the intense band at 100 cm$^{-1}$ may be assigned either to the group of bands predicted at 137.8, 138.5, and 139.2 cm$^{-1}$, or to these bands together with the peak at 107 cm$^{-1}$.
The intensity of the well defined peak predicted by DFT at $ca.$ 400 cm$^{-1}$ includes two normal modes at 403.5 and 405 cm$^{-1}$ which are associated with the torsion of the pyrazine. It undergoes a blueshift upon gas adsorption of 0.8 cm$^{-1}$ and a redshift of 2.06 cm$^{-1}$ for the duplicated mode. The predicted negligible change in the energy of this vibration by the DFT calculations is in agreement with the small change found experimentally (see Fig.~\ref{Fig1}).

\subsection{The SO$_2$ binding mechanism}
The molecular orbital interaction of SO$_2$ with this MOF was previously studied by Arc\'is-Castillo et {\sl al.}~\cite{Real2013} using DFT calculations and it was reported before by many authors for other metal complexes ~\cite{Ibers1969,Ryan1981,Mingos1978,Mingos2014}.  The results of our calculations agree with those of Arc\'is-Castillo et {\sl al.} except for the interpretation: we do not identify the mixing between the hybridized Pt d$_{z^2}$-p$_z$ orbital ~\cite{YvesJean2005} and the $\pi^*$ LUMO (lowest unoccupied molecular orbital) of the SO$_2$ as a $\pi$-backbonding  because of the absence of $\pi$ symmetry in this interaction ~\cite{YvesJean2005}. Below we discuss in more detail the molecular orbital interaction between SO$_2$ and the metal center.
\par
When SO$_2$ binds on top of a metal with square planar coordination, a strong electron donor interaction may be achieved when the $\sigma$ HOMO (highest occupied molecular orbital) of the molecule can donate electron density to the empty d$_{z^2}$ of the metal. In this case the expected binding configuration is the $\eta^1$-planar~\cite{Ryan1981} as illustrated and discussed in ref.~\citenum{Real2013}. Because for a d$^8$ electron count the d$_{z^2}$ is doubly occupied, the molecule can act only as an electron acceptor~\cite{Mingos2014}. Since the d$_{z^2}$ is fully occupied, the molecule orients in such a way to decrease the antibonding $\sigma$-type of interaction between the HOMO of the molecule and the Pt d$_z^2$, thus adopting an $\eta^1$-pyramidal configuration~\cite{Ryan1981} with the angle between the Pt and the molecular plane of the SO$_2$ being $\theta$=101.2$^{\circ}$ (PBE+D2). In this bent configuration, the LUMO of the molecule interacts with a polarized molecular orbital of Pt (see Fig.~S6) which is a hybridized Pt d$_{z^2}$-p$_z$ orbital, the polarized character resulting from the addition (bigger lobe) and cancellation (smaller lobe) of the two d$_{z^2}$ and p$_z$ amplitudes~\cite{YvesJean2005}. Such polarized character reduces the antibonding interaction.
Possibly because of the bent configuration that would reduce the $\pi$ type overlap, the interaction of the $\pi$*  LUMO does not occur with the d$_{xz}$ or d$_{yz}$ orbitals of Pt.

\vspace*{1cm}
\begin{figure}[hb!]
  \centering
 \includegraphics[scale=0.55]{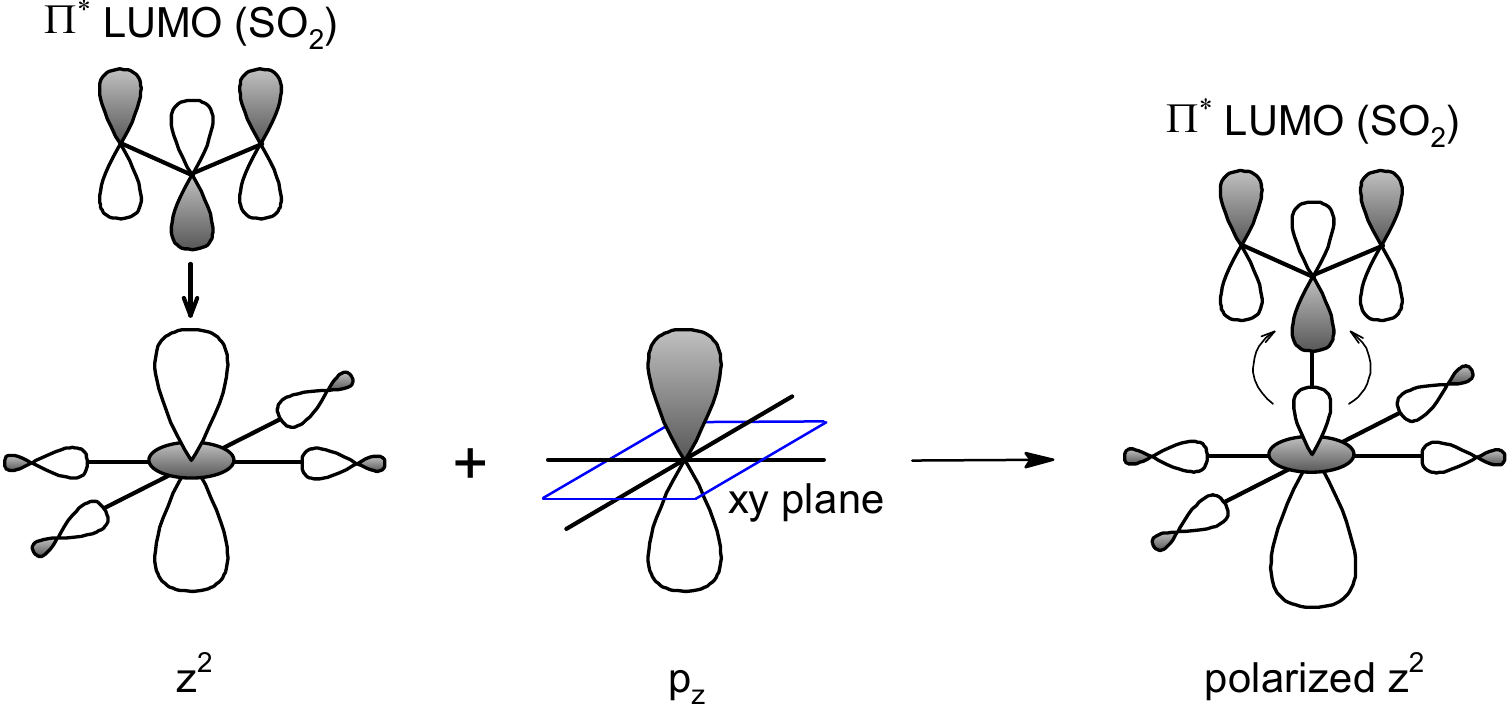}
  \vspace{0cm}
  \caption{Molecular orbital picture of the interaction between the SO$_2$ LUMO and the polarized hybridized d$_{z^2}$-p$_z$ orbital of Pt. Because this unoccupied electronic state involves a contribution from the Pt-d$_{z^2}$ that is fully occupied in the bare material, such interaction results in a metal$\to$molecule charge transfer.}
  \label{Fig6}
\end{figure}

\begin{table*}[hb!]
  \centering
  \small
\begin{tabular}{l|c|c|c|c}
\hline\hline
                   &  E$_{bind}$(eV)  & d(S-Pt) (\AA)  &  $\angle{\text{(Pt-S-O)}}$ ($\degree$)& $\angle{\text{(O-S-O)}}$ ($\degree$)\\ \hline \hline
PBE                &    0.239        & 2.895          &   102.4             & 116.8 \\ \hline
PBE+D2             &    0.769        & 2.933          &   101.2             & 117.2 \\ \hline
PBE+D2+U           &    0.683       & 3.470           &   95.2             &  118.2  \\ \hline
PBE+D2[U]           &    0.730       & 3.470           &   95.2             &  118.2  \\ \hline
PBEsol+D2          &    0.911        & 2.786          &   102.1            &  116.8 \\ \hline \hline
\end{tabular}
\caption{Comparison of the binding energy, bond distance and bond angles computed using the different functionals. The PBE+D2 computed SO$_2$ bond angle for the isolated molecule is 119.4$\degree$.}
\label{tabx}
\end{table*}

On the basis of the discussed binding mechanism, SO$_2$ is thus expected to act as an electron acceptor. 
The Bader charge analysis peformed in this work using PAW pseudopotentials reveals a rather small charge transfer from the MOF to the molecule of 0.12 electrons, suggesting that other effects possibly related to the MOF skeleton may contribute to the large value of the computed binding energy per SO$_2$ molecule, i.e. 0.769 eV (with PBE+D2). This value decreases to 0.210 eV upon full geometrical optimization after the D2 correction is removed.
We note the stronger binding computed by Arc\'is-Castillo et {\sl al.}~\cite{Real2013} using PBEsol+D2. They predict a binding energy of 0.977 eV, in agreement with the value that we compute when using the same functional, i.e. 0.953 eV.
To corroborate this, we computed the binding energy of SO$_2$ only with the two Fe[Pt(CN)$_4$] planes, respectively below and above the molecule, by fixing all the atomic coordinates to the relaxed geometry in the MOF. The lattice parameters in the plane are set to those of the MOF while we imposed a large $c$ cell parameter in order to reduce interaction between the images along $z$.  The binding energy is 0.274 eV. This large energy difference between the two cases points to a binding mechanism originating from a combined charge transfer mechanism with the Fe[Pt(CN)$_4$] planes via the SO$_2$-Pt interaction, together with a strong interaction with the rest of the MOF skeleton. \par

An efficient way to improve the treatment of electronic correlations in DFT is to adopt the DFT+U approach where the Hubbard $U$ correction is computed in terms of some localized Hubbard states. We employ the simplified rotationally invariant formulation\cite{Dudarev} as implemented in Quantum Espresso. To study how the change in hybridization~\cite{Mariano2021} of the localized d states of Pt upon $U$ correction affects the binding mechanism\cite{Cococc2016}, we employ ortho-atomic projectors to compute self-consistently the linear-response $U$ parameters associated to the $d$ states of Pt (5.2 eV) and Fe (7.5 eV), using density functional perturbation theory \cite{Timrov2018}. By setting the atomic coordinates to the optimized PBE+D2, we compute a DFT+U binding energy of 0.617 eV. When the geometry is allowed to fully relax using PBE+D2+U, we predict a SO$_2$ binding energy of 0.680 eV.
The full optimization with Hubbard $U$ gives a significantly different SO$_2$ binding geometry with respect to PBE+D2 case. Tab.~\ref{tabx} reports the binding energy, bond distance and bond angles computed with the different schemes. The S-Pt bond distance changes from 2.933 \AA\ (PBE+D2) to 3.460 \AA\ (PBE+D2+U) and the Pt-S-O angle from 101.2$\degree$ to 95.0$\degree$. The SO$_2$ bond angle, O-S-O, slighlty increases upon $U$ correction (see Tab.~\ref{tabx}). The projected density of states shown in Fig.~S7 reveals a lower contribution from the polarized Pt atoms (see Fig.~\ref{Fig6}) to the states associated with the LUMO of the molecule, consistent with a reduced charge transfer mechanism and larger SO$_2$ bending angle.  We note that the reported experimental Pt-S bond distance at 120 K is 2.585 \AA~\cite{Real2013}, significantly shorter than the PBE+D2+U one. This elongation of bond lengths upon $U$ correction has already been reported and discussed by some of us~\cite{Mariano2021} and is the reason behind the choice of the PBE+D2 geometry in the calculation of the vibrational modes. Because of this and because of the use of DFT+U may lead to a bias when computing total energy differences\cite{Lorenzo2020}, we also report the binding energy computed using a DFT+U density-corrected approach. This consists in using the PBE+D2 total energy evaluated on the PBE+D2+U electronic density, an approach that we have named PBE+D2[U]. Using the PBE+D2 geometry we obtain a value of 0.730 eV.\par
This result together with the above analysis of the binding mechanism indicate that the SO$_2$ adsorption occurs via a physisorption mechansim, the high value of the computed binding energy arising from a combination of stabilizing electrostatic and van der Walls forces.\par

\begin{table*}[hb!]
  \centering
  \small
  \begin{tabular}{c  c  c  c}
  \hline  
    Experiment (INS)  & DFT - pz@MOF & DFT - gas phase pz & Assignment \\
  \hline
    -     & 3168 3167 3165 3164  &  3059 3067  & CH stretching \\
    -     & 3155 3153 3152 3150 &    3088 3092 & CH stretching \\
    1601  & 1597 1596     & 1551 &  CH in-plane bending, CC stretch               \\
    1482  & 1487 1486     & 1525 &  CN and CC stretch                             \\
    1472  & 1459 1458     & 1472 &  CH in-plane bending, CN stretch               \\
    1430  & 1426.8 1426.3 & 1389 &  CH in-plane bending, CN and CC stretch        \\
    1338  & 1323 1322     & 1324 &  CH in-plane bending, CN stretch               \\
    1220  & 1263.9 1263.6 & 1225 &  CN and CC stretch                             \\   
    1220  & 1211 1208     & 1209 &  CH in-plane bending, CN and CC stretch        \\
    1105  & 1105 1104     & 1058 &  CH in-plane bending, CN and CC stretch        \\
    1105  & 1099 1093     & 1128 &  CH in-plane bending, CN stretch               \\
    1105  & 1047 1045     & 996  &  CN in-plane bending                            \\
    1105  & 1032 1031     & 1009 &  Breathing mode                                \\
    959   & 960 959       & 967  &  CH out-of-plane bending, CNC twisting         \\  
    959   & 949.9 949.4   & 959  &  CH out-of-plane bending, CNC wagging          \\
    876   & 862 857       & 913  &  CH out-of-plane bending, CNC twisting         \\
    817   & 820 819       & 725  &  CH out-of-plane bending, CNC wagging          \\
    749   & 735.6 735.5   & 757  &  CH out-of-plane bending, CNC wagging          \\
    696   & 687 685       & 690  &  CNC in-plane bending                          \\
    674   & 680 676       & 580  &  CNC in-plane bending                          \\
    394   & 403 404       & 312  &  CH out-of-plane bending, CNC twisting         \\
     -    &566 525& 413  &  CH out-of-plane bending, CNC wagging          \\
  \hline
  \end{tabular}
  \caption{Energy position of the peaks measured from the INS data for the bare MOF, the calculated values in the bare MOF and for pyrazine in gas phase, and their assignement. We could not find a correspondance for the modes whose energy is re}
  \label{tab1}
\end{table*}

\subsection{High-energy region}
The high energy region, which is dominated by the vibrations of the pyrazine, negligibly changes upon gas adsorption. We have computed the G($\omega$) of the pyrazine in gas phase and we report the comparison with the partial G($\omega$) of the pyrazine in the compound in Fig.~S8. The complexation of the pyrazine in the MOF slightly modifies the position of the vibrational modes with respect to the gas phase. A comparison and a brief description of the modes is reported in Table ~\ref{tab1}. These were studied by various authors by employing IR and Raman spectroscopies~\cite{Ito1956,Lord1957,Katritzky1959,Califano1964,Simmons1964,Arenas1985,Endredi2003} and INS~\cite{Kearly1996}. Three different spectral regions can be identified for the free and complexed pyrazine: (i) in-plane CH bendings and ring stretching motions in [990 - 1600] cm$^{-1}$, (ii) intense bands associated with the CH out-of-plane deformation vibrations at [700 - 990] cm$^{-1}$, and (iii) CH stretching near 3000 cm$^{-1}$.  The vibrations with predominant CC and CN character can be further divided into C-C and C-N stretching modes at 1058 - 1551 cm$^{-1}$, in-plane bending modes at 580 cm$^{-1}$ and 690 cm$^{-1}$ and out-of-plane bending in the region 750 - 990 cm$^{-1}$ and bellow 500 cm$^{-1}$ like 413 cm$^{-1}$ and 312 cm$^{-1}$. The characteristic breathing mode of the pyrazine appears at 1009  cm$^{-1}$. As a general trend the complexation with Fe(II)  blueshifts the vibrational modes of the free pyrazine. This effect is possibly driven by an electron-acceptor effect of the pyrazine from the occupied d orbitals of the metal~\cite{Toma1972} that increases the $\pi$ cloud of the ring: thus the stronger double bond character increases the vibrational frequencies \cite{Thompson2018}. We note that the CH out-of-plane bending mode at 913 cm$^{-1}$ undergoes the opposite effect and redshifts to $ca.$ 859 cm$^{-1}$ thus appearing in the gap between [760 - 910] cm$^{-1}$ of the free pyrazine (see Fig.~\ref{Fig2}).

\subsection{Choice of the supercell}
\begin{figure}[ht!]
  \includegraphics[scale=0.45]{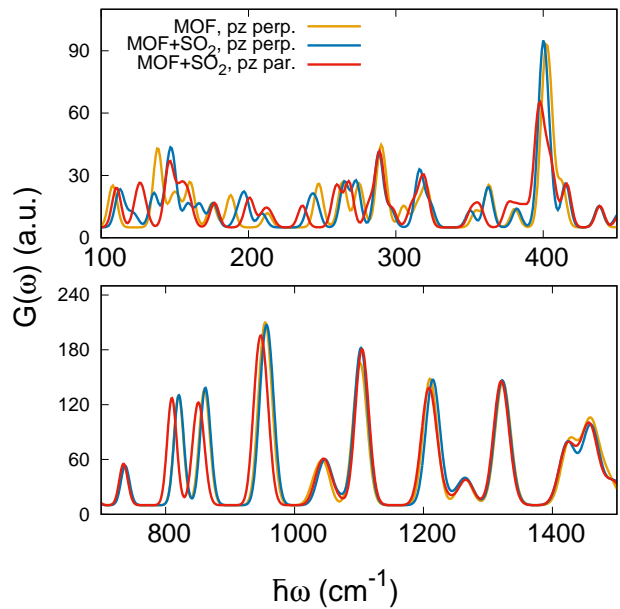}
  \caption{Total G($\omega$) in the low (upper panel) energy and high energy (lower panel) computed for the bare material with perpendicularly oriented pyrazines (yellow), and for the material loaded with one SO$_2$ per Pt site computed by employing perpendicular (blue) or parallel (red) pyrazines.}
  \label{Figs}
\end{figure}

Several authors have reported a specific long range ordering of the pyrazine ligands within these materials as a consecuence of gas adsorption~\cite{Ohba2009,Kepert2009,Aravena2014}. In the absence of diffraction data, we assessed the ordering of the pyrazine upon SO$_2$ adsorption using DFT calculations. We performed two supercell calculations where we impose that the pyrazines are oriented either perpendicularly with neighboring ones or parallel. We find negligible energy differences between these two configurations when the material is loaded with one SO$_2$ molecule per Pt atom. We recall that for the bare material such energy difference becomes 39 meV/f.u. ~\cite{Fernandez-Blanco2021}. The comparison between the corresponding generalized phonon density of states and that of the bare material, however, shows that the configuration with all pyrazines oriented parallel with one another results in a predicted G($\omega$) exhibiting some deviations from the bare material that are not observed experimentally. The comparison between the empty (pyrazine perpendicular) and the loaded material is reported in Fig.~\ref{Figs}.

The configuration with parallel pyrazines upon SO$_2$ adsorption results in a noticeable change in the intense peak at 400 cm$^{-1}$ with respect to the bare material, a feature that is not observed in the experiment. Similarly, the high energy features between 800 and 1000 cm$^{-1}$ redshift in the configuration with parallel pyrazines, suggesting that the INS data may correspond to a configuration with pyrazines oriented perpendicularly with each other both before and after gas adsorption.

\par We tested a few functional choices on the computed G($\omega$) for the bare material. For this we employed the primitive cell (thus imposing parallel pyrazines) and the exchange and correlation functionals PBEsol+D2, rev-vdW-DF2, and PBE+D2. The three functionals give very similar results. At high energy, the spectra negligibly depend on the specific choice while at low energy, as expected, they are more strongly affected by the approximation to the exchange and correlation energy~\cite{Koi2018}, as shown in Fig.~S9. Specifically, we see a tendency to predict higher frequencies in this region using PBEsol+D2, as compared with PBE+D2, consistent with shorter bond distances and smaller lattice parameters.

\section{Conclusion}
The SO$_2$ adsorption mechanism in the Fe(pz)[Pt(CN)$_4$] Hofmann clathrate is here probed by combining inelastic neutron scattering experiments and DFT calculations. The most noticeable change upon adsorption is the blueshift of the peak measured at $ca.$ 100 cm$^{-1}$ which we attribute to an increase of the cyanide out-of-plane bending force constant due to steric hindrance and to the hindered rotation of the pyrazine around the $z$ axis. The experimental observations are consistent with an adsorption mechanism being a combined charge transfer, electrostatic and dispersion interactions, as predicted by the DFT calculations. The high-energy peaks which are associated with the internal vibrations of the pyrazine molecule are negligibly affected by the presence of the physisorbed gas molecule. Based on these findings we expect a different signature of the adsorption of different molecules such as CO or CO$_2$ on the INS data and are confident that the present joint computational and experimental scheme will allow to characterize the gas interaction mechanism in these family of materials.

\begin{acknowledgement}
This work benefited from the support of the project ANR-15-CE06-0003-01 funded by the French National Agency for Research. Calculations were performed using resources granted by GENCI under the CINES and TGCC grant numbers A0020907211 and A0110907211. Additionally, the froggy and Dahu platform of the CIMENT infrastructure, which is supported by the Rhone-Alpes region (GRANT CPER07\underline{ }13 CIRA) and the Equip@Meso project, was employed for the calculations. We thank the ILL for the PhD contract of A.F.B. and for the beamtime allocation under experiment numbers 7-02-108, 7-05-513, and 7-05-430 (doi.ill.fr/10.5291/ILL-DATA.7-05-430). J.A.R.~acknowledges resources from Grant PID2019-106147GB-I00 funded by MCIN/AEI/10.13039/501100011033.
\end{acknowledgement}

\bibliography{bibtex}
\end{document}